\documentclass[pre,aps]{revtex4-2}
\usepackage{mathpazo}
\usepackage[T1]{fontenc}
\usepackage[latin9]{inputenc}
\usepackage{color}
\usepackage{amsmath}
\usepackage{amssymb}
\usepackage{graphicx}
\usepackage{subfigure}
\usepackage{ulem}

\begin{document}

\title{Stable patterns in the Lugiato-Lefever equation with a confined
vortex pump}
\author{Shatrughna Kumar$^{1}$}
\author{Wesley B. Cardoso$^{2}$}
\author{Boris A. Malomed$^{1,3}$}
\affiliation{$^{1}$Department of Physical Electronics, School of Electrical Engineering,
Faculty of Engineering, and Center for Light-Matter Interaction, Tel
Aviv University, Tel Aviv 69978, Israel}
\affiliation{$^{2}$Instituto de Física, Universidade Federal de Goiás, 74.690-900, Goiânia,
Goiás, Brazil}
\affiliation{$^{3}$Instituto de Alta Investigación, Universidad de Tarapacá, Casilla
7D, Arica, Chile}

\begin{abstract}
We introduce a model of a passive optical cavity based on a novel variety of
the two-dimensional Lugiato-Lefever equation, with a localized pump carrying
intrinsic vorticity $S$, and the cubic or cubic-quintic nonlinearity. Up to $%
S=5$, stable confined vortex-ring states (vortex pixels) are produced by
means of a variational approximation and in a numerical form. Surprisingly,
vast stability areas of the vortex states are found, for both the
self-focusing and defocusing signs of the nonlinearity, in the plane of the
pump and loss parameters. When the vortex-rings are unstable, they are
destroyed by azimuthal perturbations which break the axial symmetry. The
results suggest new possibilities for mode manipulations in passive
nonlinear photonic media by means of appropriately designed pump beams.
\end{abstract}

\maketitle

\section{Introduction and the model}

Optical solitons are a broad class of self-trapped states maintained by the
interplay of nonlinearity and dispersion or diffraction in diverse photonic
media \cite{Kivshar_03,book}. In addition to that, dissipative optical
solitons are supported by the equilibrium of loss and gain or pump,
concomitant to the nonlinearity-dispersion/diffraction balance \cite%
{Rosanov,Ferreira}. Dissipative solitons have been studied in detail,
theoretically and experimentally, in active setups, with the loss
compensated by local gain (essentially, provided by lasing), being modelled
by one- and two-dimensional (1D and 2D) equations of the complex
Ginzburg-Landau (CGL) type \cite{Grelu,2D}.

In passive nonlinear optical cavities, the losses are balanced by the pump
field supplied by external laser beams, with the appropriate models provided
by the Lugiato-Lefever (LL) equations \cite{Lugiato_PRL87}. This setting was
also studied in the 1D and 2D forms \cite%
{Tlidi_C17,Panajotov_EPJB17,Tlidi_AOP22,Wabnitz}. Widely applied in
nonlinear optics, equations of the LL type play a crucial role in
understanding fundamental phenomena such as the modulation instability (MI)
and pattern formation in dissipative environments \cite{Tlidi_C17}-\cite%
{Dong_PRR21}.
The relevance of these models extends to the exploration of complex dynamics
of various nonlinear photonic modes, tremendously important applications
being the generation of Kerr solitons and frequency combs in passive
cavities \cite{Valcarcel_PRA13}-\cite{Dong_PRR21}, as well as the generation
of terahertz radiation \cite{Huang_PRX17}. In addition to rectilinear cavity
resonators, circular ones can be used too \cite{Kartashov_OE17}. In many
cases, they operate in the whispering-gallery regime \cite{Coillet_PJ13}-%
\cite{Cao_CSF23}.

In most cases, solutions of the one- and two-dimensional LL equations are
looked for under the action of the spatially uniform pump, which
approximately corresponds to the usual experimental setup. However, the use
of localized (focused) pump beams is possible too, which makes it relevant
to consider LL equations with the respective shape of the pump terms. In
fact, truly localized optical modes in the cavities can be created only in
this case, otherwise the uniform pump supports nonzero background of the
optical field. In particular, exact analytical solutions of the LL equations
with the 1D pump represented by the delta-function, and approximate
solutions maintained by the 2D pump in the form of a Gaussian were reported\
in Ref. \cite{Cardoso_SR17}. In Ref. \cite{Kartashov_OE17}, the LL equation
for the ring resonator with localized pump and loss terms produced nonlinear
resonances leading to multistability of nonlinear modes and coexisting
solitons, that are associated with spectrally distinct frequency combs.

Further, solutions for fully localized robust pixels with zero background
were produced by the 2D LL equation, incorporating the spatially uniform
pump, self-focusing or defocusing cubic nonlinearity, and a tight confining
harmonic-oscillator potential \cite{Cardoso_EPJD17}. Additionally, this
model with a vorticity-carrying pump gives rise to stable vortex pixels. In
particular, in the case of the self-defocusing sign of the nonlinear term,
the pixels with zero vorticity and ones with vorticity $S=1$ were predicted
analytically by means of the Thomas-Fermi (TF) approximation.

In this work, we introduce the 2D LL equation for complex amplitude field $%
u\left( x,y,t\right) $ of the light field, with the cubic or cubic-quintic
nonlinearity:%
\begin{equation}
\frac{\partial u}{\partial t}=-\alpha u+\frac{i}{2}\nabla ^{2}u+i\sigma
\left( \left\vert u\right\vert ^{2}-\eta ^{2}\right) u-ig\left( \left\vert
u\right\vert ^{4}-\eta ^{4}\right) u+f(r)e^{iS\theta },  \label{LLE}
\end{equation}%
and a confined pump beam represented by factor $f(r)$. Here $i$ is the
imaginary unit, $\alpha >0$ is the loss parameter, $\sigma =+1$ and $-1$
corresponds, respectively, to the self-focusing and defocusing Kerr (cubic)
nonlinearity, $g>0$ or $g<0$ represents the self-defocusing or focusing
quintic nonlinearity (which often occurs in optical media \cite%
{Spain,Boudebs,Cid}, in addition to the cubic term), parameter $\eta $
defines the cavity mismatch, which is $\sigma \eta ^{2}-g\eta ^{4}$ (the
coefficient multiplying the linear term $\sim iu$), in terms of the
linearized LL equation, and
\begin{equation}
f(r)=f_{0}r^{S}\exp \left( -r^{2}/W^{2}\right)  \label{f}
\end{equation}%
written in terms of polar coordinates $\left( r,\theta \right) $,
corresponds to the confined pump beam with real amplitude $f_{0}$, radial
width $W$, and integer vorticity $S\geq 1$. Vortex beams, shaped by the
passage of the usual laser beam through an appropriate phase mask, are
available in the experiment \cite{Cid}.

Equation (\ref{LLE}) is written in the scaled form. All figures are plotted
below in the same notation. In physical units, $r=1$ and $t=1$ normally
correspond to $\symbol{126}50$\textrm{\ }$\mathrm{\mu }$m and $\sim 50$ ps,
respectively. Then, the typical width $W=2$, considered below, corresponds
to the pump beam with diameter $\sim 100$\textrm{\ }$\mathrm{\mu }$m, which
an experimentally relevant value. Accordingly, the characteristic evolution
time in simulations presented below, $\grave{t}\sim 100$, corresponds to
times $\symbol{126}5$ ns.

Stationary solutions of Eq. (\ref{LLE}) are characterized by values of the
total power (alias norm),
\begin{equation}
P=\int_{-\infty }^{+\infty }dx\int_{-\infty }^{+\infty }dy~\left\vert
u\left( x,y\right) \right\vert ^{2}\equiv 2\pi \int_{0}^{\infty }\left\vert
u(r,t)\right\vert ^{2}rdr,  \label{P}
\end{equation}%
and angular momentum,
\begin{equation}
M=i\int_{-\infty }^{+\infty }dx\int_{-\infty }^{+\infty }dyu^{\ast }\left(
y\partial _{x}u-x\partial _{y}u\right) dxdy  \label{M}
\end{equation}%
(with $\ast $ standing for the complex conjugate), even if the power and
angular momentum are not dynamical invariants of the dissipative equation (%
\ref{LLE}). In the case of the axisymmetric solutions with vorticity $S$,
i.e., $u\left( x,y\right) =u(r)e^{iS\theta }$ \cite{Andrews,Ramaniuk,PhysD},
the expressions for the power and angular momentum are simplified:
\begin{equation}
P=2\pi \int_{0}^{\infty }\left\vert u(r)\right\vert ^{2}rdr,~M=SP.
\label{PM}
\end{equation}

Our objective is to construct \emph{stable} ring-shaped vortex solitons
(representing vortex pixels, in terms of plausible applications), as
localized solutions of Eq. (\ref{LLE}) with the same $S$ as in the pump term
(\ref{f}). The stability is a challenging problem, as it is well known from
the work with models based on the nonlinear Schrödinger and CGL equations
that (in the absence of a tight confining potential) vortex-ring solitons
are normally vulnerable to the splitting instability. In the case of a
narrow ring shape, the splitting instability may be considered as
quasi-one-dimensional MI of the ring against azimuthal perturbations which
break its axial symmetry \cite{PhysD,book}. The azimuthal MI is driven by
the self-focusing nonlinearity, and inhibited by the self-defocusing.

To produce stationary solutions for the vortex solitons in an approximate
analytical form (parallel to the numerical solution), we employ a
variational approximation (VA). Our results identify regions of the
existence and stability of the vortex solitons with $1\leq S\leq 4$ in the
space of parameters of Eqs. (\ref{LLE}) and (\ref{f}) (in particular, in the
plane of $\left( f_{0},\alpha \right) $) for both signs of the cubic
nonlinearity, $\sigma =\pm 1$, while the mismatch parameter is fixed to be $%
\eta =1$ by dint of scaling. The stability areas are vast, provided that the
loss coefficient $\alpha $ is, roughly speaking, not too small. A majority
of the results are produced for the pure cubic model, with $g=0$, but the
effect of the quintic term, with $g\neq 0$, is considered too. Quite
surprisingly, a stability area for the vortices with $S\leq 3$ is found even
in the case of $\sigma =+1,g<0$, when both the cubic and quintic terms are
self-focusing, which usually implies strong propensity to the azimuthal
instability of the vortex rings \cite{PhysD}.

The rest of the paper is structured as follows. The analytical approach,
based on the appropriate VA, is presented in Sec. \ref%
{sec:Analytical-estimates}. An asymptotic expression for the tail of the
vortex solitons, decaying at $r\rightarrow \infty $, is found too in that
section. Systematically produced numerical results for the shape and
stability of the vortex solitons are collected (and compared to the VA
predictions) in Sec. \ref{sec:Numerical-results}. The paper is concluded by
Sec. \ref{sec:Conclusion}.

\section{Analytical considerations \label{sec:Analytical-estimates}}

This section summarizes the analytical part of the work and results produced
by this part. Two directions of the analytical considerations for the
present model are possible: the investigation of the decaying
\textquotedblleft tails" of the localized stationary states, in the
framework of the linearized model, and detailed development of VA for the
full model, including the nonlinear terms.

\subsection{Asymptotic forms of the vortex solitons}

Direct consideration of the linearized version of Eqs. (\ref{LLE}) and (\ref%
{f}) readily produces an explicit result for the soliton's tail decaying at $%
r\rightarrow \infty $:%
\begin{equation}
u(r,\theta )\approx (i/2)W^{4}r^{S-2}\exp \left( -r^{2}/W^{2}+iS\theta
\right) ,  \label{r to infinity}
\end{equation}%
with the power of the pre-Gaussian factor, $r^{S-2}$, which is\ lower than
that in the pump term, $r^{S}$. Due to this feature, the asymptotic
expression (\ref{r to infinity}) formally predicts a maximum of local power $%
\left\vert u(r)\right\vert ^{2}$ at $S>2$, at
\begin{equation}
r^{2}=r_{\max }^{2}\equiv \sqrt{(S/2-1)}W.  \label{max1}
\end{equation}%
A local maximum is indeed observed in numerically found radial profiles of
all vortex solitons (see Figs. \ref{F2}, \ref{FS4_f1_sigma=-1}, \ref{F6}(b), %
\ref{F8}(b), \ref{F9}, and \ref{F10} below, for $S=1$, $4$, $2$, $2$, $1$,
and $3$, respectively). In fact, for these cases Eq. (\ref{max1}) predicts
values of $r_{\max }$ which are smaller by a factor $\simeq 0.6$ than the
actually observed positions of the maxima. The discrepancy is explained by
the fact that the asymptotic expression (\ref{r to infinity}) is valid at
values of $r$ which are essentially larger than $r_{\max }$.

In a looser form, one can try to construct an asymptotic approximation for
the solution at moderately large $r$, by adopting the \textit{ansatz} which
follows the functional form of the pump term (\ref{f}), \textit{viz}.,%
\begin{equation}
u(r,\theta )\approx \mathrm{u}(r)r^{S}\exp \left( -r^{2}/W^{2}+iS\theta
\right) ,  \label{ansatz0}
\end{equation}%
where $\mathrm{u}(r)$ is a complex slowly varying function, in comparison
with those which are explicitly present in \textit{ansatz} (\ref{ansatz0}).
Substituting the \textit{ansatz} in Eq. (\ref{LLE}) and omitting derivatives
of the slowly varying function, one can develop an approach which is akin to
the TF approximation applied to the model with the tight trapping potential
in Ref. \cite{Cardoso_SR17}. The result for the linearized version of Eq. (%
\ref{LLE}), which implies a small amplitude of the mode pinned to the pump
beam, is%
\begin{equation}
\mathrm{u}(r)=f_{0}\left[ \alpha -i\left( \frac{2r^{2}}{W^{4}}-\frac{2(S+1)}{%
W^{2}}-\sigma +g\right) \right] ^{-1},  \label{TF}
\end{equation}%
where, as said above, $\eta =1$ is substituted. In the limit of $%
r\rightarrow \infty $, Eqs. (\ref{TF}) and (\ref{ansatz0}) carry over into
the asymptotically rigorous expression (\ref{r to infinity}). On the other
hand, Eq. (\ref{TF}) predicts a maximum of the local power at%
\begin{equation}
r^{2}=\left( r_{\max }^{2}\right) _{\mathrm{TF}}=\left( S+1\right) W^{2}+%
\frac{\sigma -g}{2}W^{4},  \label{max2}
\end{equation}%
cf. Eq. (\ref{max1}). One may expect that the prediction of the local
maximum of the vortex soliton at point (\ref{max2}) is valid when it yields
values of $\left( r_{\max }^{2}\right) _{\mathrm{TF}}$ which are large
enough, i.e., if $S$ and $W$ are relatively large. Indeed, the comparison
with the numerically produced profiles of the vortex solitons, displayed
below in Figs. \ref{F5} and \ref{F9}, demonstrates that Eq. (\ref{max2})
predicts, relatively accurately, $\left( r_{\max }\right) _{\mathrm{TF}}=4$
for $S=5$, $W=2$, $\sigma =-1$, and $g=0$. However, the prediction given by
Eq. (\ref{max2}) is not accurate for $S=1$ and $2$.

Finally, in the limit of $r\rightarrow 0$, the asymptotic form of the
solution is simple, $u(r,\theta )\approx u_{0}r^{S}$, but constant $u_{0}$\
cannot be found explicitly, as it depends on the global structure of the
vortex-soliton solution. In particular, the crude TF approximation given by
Eq. (\ref{TF}) yields $u_{0}=f_{0}\left[ \alpha +i\left( \frac{2(S+1)}{W^{2}}%
+\sigma -g\right) \right] ^{-1}$.

\subsection{The variational approximation (VA)}

A consistent global analytical fit for the vortex solitons may be provided
by\ VA, based on the Lagrangian of the underlying equation \cite{book}.
While Eq. (\ref{LLE}), which includes the linear dissipative term, does not
have a Lagrangian structure, it can be converted into an appropriate form by
the substitution suggested by Ref. \cite{Manakov}, which absorbs the
dissipative term:
\begin{equation}
u\left( r,\theta ,t\right) =U(r,t)e^{iS\theta -\alpha t},  \label{uU}
\end{equation}%
producing the following time-dependent equation for complex function $U(r,t)$%
, where, as said above, we set $\eta =1$ by means of scaling:
\begin{equation}
\frac{\partial U}{\partial t}=\frac{i}{2}\left( \frac{\partial ^{2}}{%
\partial r^{2}}+\frac{1}{r}\frac{\partial }{\partial r}-\frac{S^{2}}{r^{2}}%
\right) U+i\sigma \left( \left\vert U\right\vert ^{2}e^{-2\alpha t}-1\right)
U-ig\left( \left\vert U\right\vert ^{4}e^{-4\alpha t}-1\right)
U+f(r)e^{\alpha t}.  \label{eq-U}
\end{equation}

The real Lagrangian which precisely produces the time-dependent equation (%
\ref{eq-U}) is
\begin{gather}
L=\int_{0}^{\infty }\left[ \frac{i}{2}\left( \frac{\partial U^{\ast }}{%
\partial t}U-U^{\ast }\frac{\partial U}{\partial t}\right) +\frac{1}{2}%
\left\vert \frac{\partial U}{\partial r}\right\vert ^{2}+\left( \frac{S^{2}}{%
2r^{2}}+\sigma -g\right) |U|^{2}\right.  \notag \\
-\left. \frac{\sigma }{2}e^{-2\alpha t}|U|^{4}+\frac{g}{3}e^{-4\alpha
t}|U|^{6}+if(r)e^{\alpha t}\left( U^{\ast }-U\right) \right] rdr.  \label{L}
\end{gather}%
The simplest \textit{ansatz}\ which may be used as the basis for VA follows
the form of the pump term (\ref{f}):
\begin{equation}
U(r,t)=U_{0}e^{i\phi }r^{S}\exp \left( -\frac{r^{2}}{W^{2}}+\alpha t\right) ,
\label{U}
\end{equation}%
where variational parameters $U_{0}$ and $\phi $ are the real amplitude and
phase shift of the solution with respect to the pump. Power (\ref{P}) for
this \textit{ansatz} is
\begin{equation}
P_{S}=\pi \Gamma (S+1)\left( \frac{W^{2}}{2}\right) ^{S+1}U_{0}^{2},
\label{PS}
\end{equation}%
where $\Gamma (S+1)\equiv S!$ is the Gamma-function, and the time-dependent
factors, $\exp \left( \pm \alpha t\right) $, mutually cancel when relations (%
\ref{uU}) and (\ref{U}) are substituted in expression (\ref{P}). Note that
the local power $|U(r)|^{2}$, corresponding to \textit{ansatz} (\ref{U}),
attains it maximum at $r^{2}=SW^{2}/2$.

The substitution of \textit{ansatz} (\ref{U}) in Lagrangian (\ref{L}) and
straightforward integration yields the respective VA Lagrangian,
\begin{eqnarray}
L_{\mathrm{VA}} &=&U_{0}e^{2\alpha t}\left\{ \left[ 6^{-(3S+2)}\Gamma \left(
3S+1\right) gW^{2(3S+1)}\right] U_{0}^{5}-\left[ 2^{-4(S+1)}\Gamma \left(
2S+1\right) \sigma W^{2(2S+1)}\right] U_{0}^{3}\right.  \notag \\
&+&\left. 2^{-(S+2)}\Gamma \left( S+1\right) W^{2S}\left[ \left( \frac{d\phi
}{dt}+(\sigma -g)\right) W^{2}+\left( S+1\right) \right] U_{0}\right.  \notag
\\
&+&\left. 2^{-(S+1)}\Gamma \left( S+1\right) W^{2(S+1)}f_{0}\sin \phi
\right\} .  \label{Leff}
\end{eqnarray}%
Then, the Euler-Lagrange equations for $U_{0}$ and $\phi $ are obtained as
\begin{equation}
\partial L_{\mathrm{VA}}/\partial U_{0}=\delta L_{\mathrm{VA}}/\delta \phi
=0,  \label{EL}
\end{equation}%
where $\delta /\delta \phi $ stands for the variational derivative. Taking
into regard that Lagrangian (\ref{Leff}) must be substituted in the
respective action, $\int L_{\mathrm{VA}}dt$, and then the action must be
actually subjected to the variation, one should apply the time
differentiation to factor $e^{2\alpha t}$ in Lagrangian (\ref{Leff}), while
deducing the appropriate form of $\delta L_{\mathrm{VA}}/\delta \phi $ in
Eq. (\ref{EL}). Once the Euler-Lagrange equations (\ref{EL}) have been
derived , we consider their stationary (fixed-point) solutions by setting $%
dU_{0}/dt=d\phi /dt=0$, which yields
\begin{equation}
U_{0}\alpha -f_{0}\cos (\phi )=0,  \label{EL1}
\end{equation}%
\begin{eqnarray}
6^{-(3S+1)}\Gamma \left( 3S+1\right)
gW^{2(2S+1)}U_{0}^{5}-2^{-2(2S+1)}\Gamma \left( 2S+1\right) \sigma
W^{2S+2}U_{0}^{3} &&+  \notag \\
+2^{-(S+1)}\Gamma \left( S+1\right) \left\{ f_{0}W^{2}\sin \phi +\left[
(\sigma -g)W^{2}+\left( S+1\right) \right] U_{0}\right\} &=&0.  \label{EL2}
\end{eqnarray}

It is relevant to mention that the evolution \ equation for the power (\ref%
{P}), which follows from Eq. (\ref{eq-U}), is%
\begin{equation}
\frac{dP}{dt}=-2\alpha P+4\pi \int_{0}^{\infty }f(r)\text{Re}\left\{
U(r,t)\right\} e^{-\alpha t}rdr.  \label{balance}
\end{equation}%
The stationary states must satisfy the balance condition, $dP/dt=0$. Then,
the substitution of \emph{ansatz} (\ref{U}) and expression (\ref{f}) for the
pump in this condition yields a simple relation,
\begin{equation}
\cos \phi =\frac{\alpha U_{0}}{f_{0}},  \label{=00003D0}
\end{equation}%
which is identical to Eq. (\ref{EL1}). In particular, Eq. (\ref{=00003D0})
implies that, for the fixed pump's amplitude $f_{0}$, the amplitude of the
established localized pattern cannot exceed the maximum value, which
corresponds to $\phi =0$ in Eq. (\ref{=00003D0}):
\begin{equation}
U_{0}\leq \left( U_{0}\right) _{\max }=f_{0}/\alpha .  \label{max}
\end{equation}

\subsection{VA for the cubic ($g=0$) and quintic ($g\rightarrow \infty $)
models}

First, we aim to predict stationary states, as solutions of Eqs. (\ref{EL1})
and (\ref{EL2}), for the model with cubic-only nonlinearity, i.e., $g=0$,
under the assumption that the loss and pump terms in Eq. (\ref{LLE}) may be
considered as small perturbations. In the lowest approximation, i.e.,
dropping the small term $\sim f_{0}$\ in Eq. (\ref{EL2}), one obtains a
relatively simple expression which predicts the squared amplitude of the
vortex soliton,
\begin{equation}
\left( U_{0}^{2}\right) _{\mathrm{VA}}^{(g=0)}=\frac{2^{2S+1}}{(2S-1)!!}%
W^{-2S}\left[ 1+\sigma \left( S+1\right) W^{-2}\right] ,  \label{U2VA}
\end{equation}%
the respective power (\ref{PS}) of the underlying \textit{ansatz} (\ref{U})
being
\begin{equation}
P_{\mathrm{VA}}^{(g=0)}=\frac{2^{S}S!\pi }{(2S-1)!!}\left[ W^{2}+\sigma
\left( S+1\right) \right] .  \label{P_cubic}
\end{equation}%
Note that expressions (\ref{U2VA}) and (\ref{P_cubic}) are always meaningful
for the self-focusing sign of the cubic nonlinearity, $\sigma =+1$, while in
the case of defocusing, $\sigma =-1$, the expressions are meaningful if they
are positive, which imposes a\ restriction on the width of the Gaussian
pump: it must be broad enough, \textit{viz}.,
\begin{equation}
W^{2}>S+1.  \label{SD_rest}
\end{equation}

\begin{figure}[tb]
\centering \includegraphics[width=0.48\columnwidth]{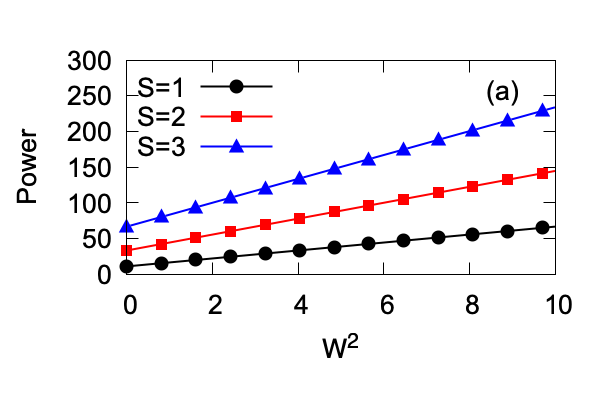} %
\includegraphics[width=0.48\columnwidth]{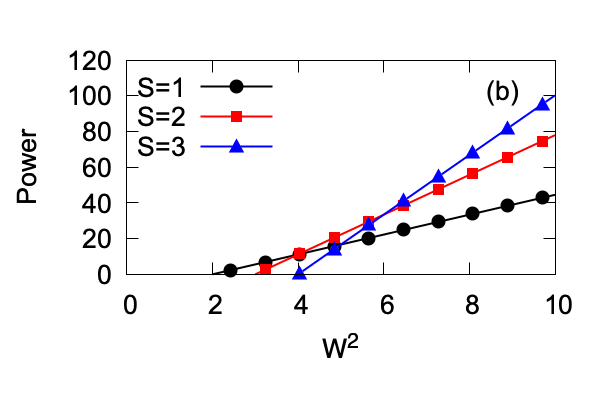}
\caption{The power of the VA solution, in the case of $g=0$ (no quintic
nonlinearity), vs. the pump's squared width, as given by Eq. (\protect\ref%
{P_cubic}), which neglects weak effects of\ the pump\ and loss (small $f_{0}$
and $\protect\alpha $), for the self-focusing ($\protect\sigma =+1$) in (a)
and defocusing ($\protect\sigma =-1$) in (b) signs of the cubic term. The
black curve with circles, the red one with squares, and the blue one with
triangles pertain to vorticities $S=1$, $2$, and $3$, respectively.}
\label{F1}
\end{figure}

The dependence of the power given by Eq. (\ref{P_cubic}) on the pump's
squared width $W^{2}$ for three different values of the vorticity, $S=1,2,3$%
, is plotted in Figs. \ref{F1}(a) and (b) for the self-focusing and
defocusing signs of the cubic term, i.e., $\sigma =+1$ and $-1$,
respectively. Note that in Fig. \ref{F1}(b) for $\sigma =-1$, there is no
solution in the region in which condition (\ref{SD_rest}) does not hold,
hence the VA solution does not exist.

In the limit of the dominant quintic nonlinearity, i.e., $g\rightarrow \pm
\infty $, opposite to the pure cubic model considered above, the asymptotic
solution of Eq. (\ref{EL2}) is%
\begin{equation}
\left( U_{0}^{2}\right) _{\mathrm{VA}}^{(g\rightarrow \pm \infty )}\approx
2^{S}\sqrt{3^{3S+1}\frac{S!}{(3S)!}}W^{-2S}  \label{Ug}
\end{equation}%
(in which the large coefficient $g$ cancels out), the respective expression
for power (\ref{PS}) being%
\begin{equation}
P_{\mathrm{VA}}^{(g\rightarrow \pm \infty )}=\frac{3^{(3S+1)/2}\pi \left(
S!\right) ^{3/2}}{2\sqrt{(3S)!}}W^{2},  \label{Pg}
\end{equation}%
cf. Eqs. (\ref{U2VA}) and (\ref{P_cubic}).

In the following section, we report results of numerical solution of Eq. (%
\ref{LLE}), comparing them to solutions of the full system of the VA
equations (\ref{EL1}) and (\ref{EL2}), which include effects of the pump and
loss terms.

\section{Numerical results \label{sec:Numerical-results}}

Simulations of Eq. (\ref{LLE}) were conducted by means of the split-step
pseudo-spectral algorithm. The solution procedure started from the zero
input, and was running until convergence to an apparently stable stationary
profile (if this outcome of the evolution was possible). This profile was
then compared to its VA counterpart, produced by a numerical solution of
Eqs. (\ref{EL1}) and (\ref{EL2}) with the same values of parameters $\alpha $%
, $\sigma =\pm 1$, $g$, and $f_{0}$, $W$, $S$ (see Eq. (\ref{f})). The
results are presented below by varying, severally, loss $\alpha $, vorticity
$S$, the pump's width $W$ and strength $f_{0}$, and, eventually, the quintic
coefficient $g$. The findings are eventually summarized in the form of
stability charts plotted in Fig. \ref{F11}.

\subsection{Variation of the loss parameter $\protect\alpha $}

\begin{figure}[tb]
\centering \includegraphics[width=0.48\columnwidth]{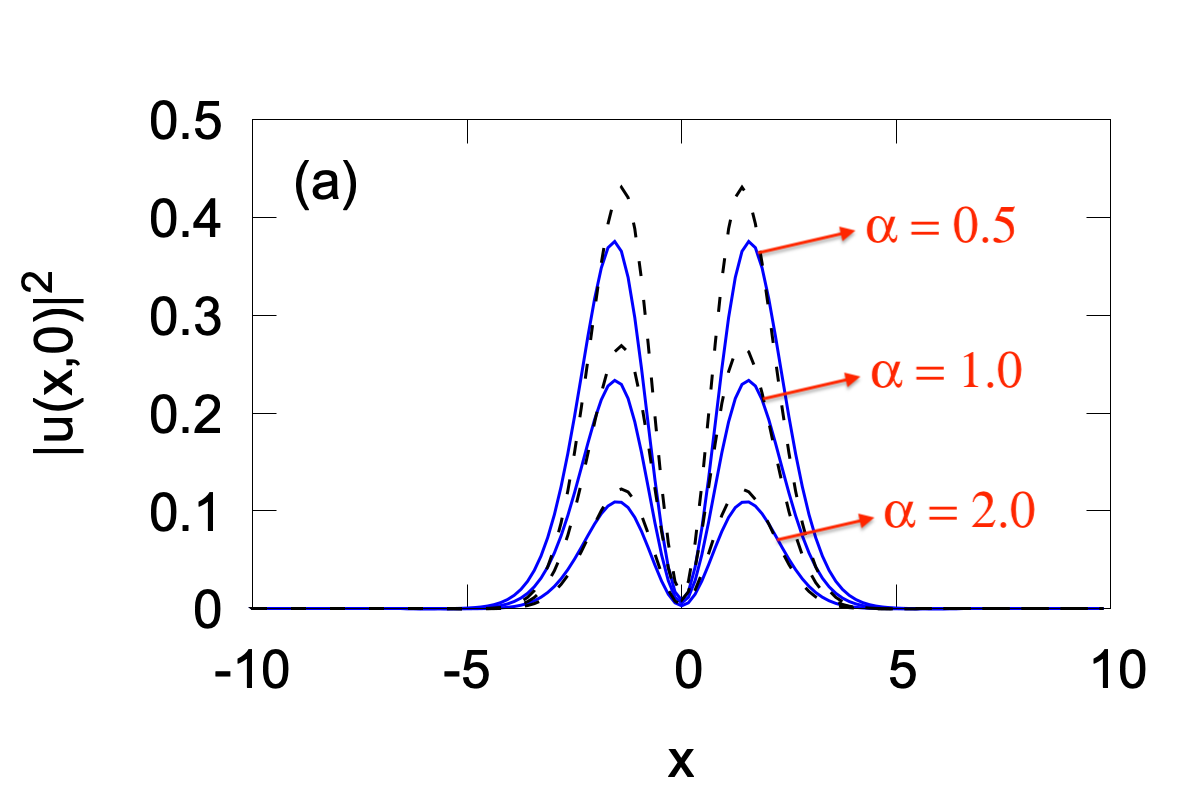} %
\includegraphics[width=0.48\columnwidth]{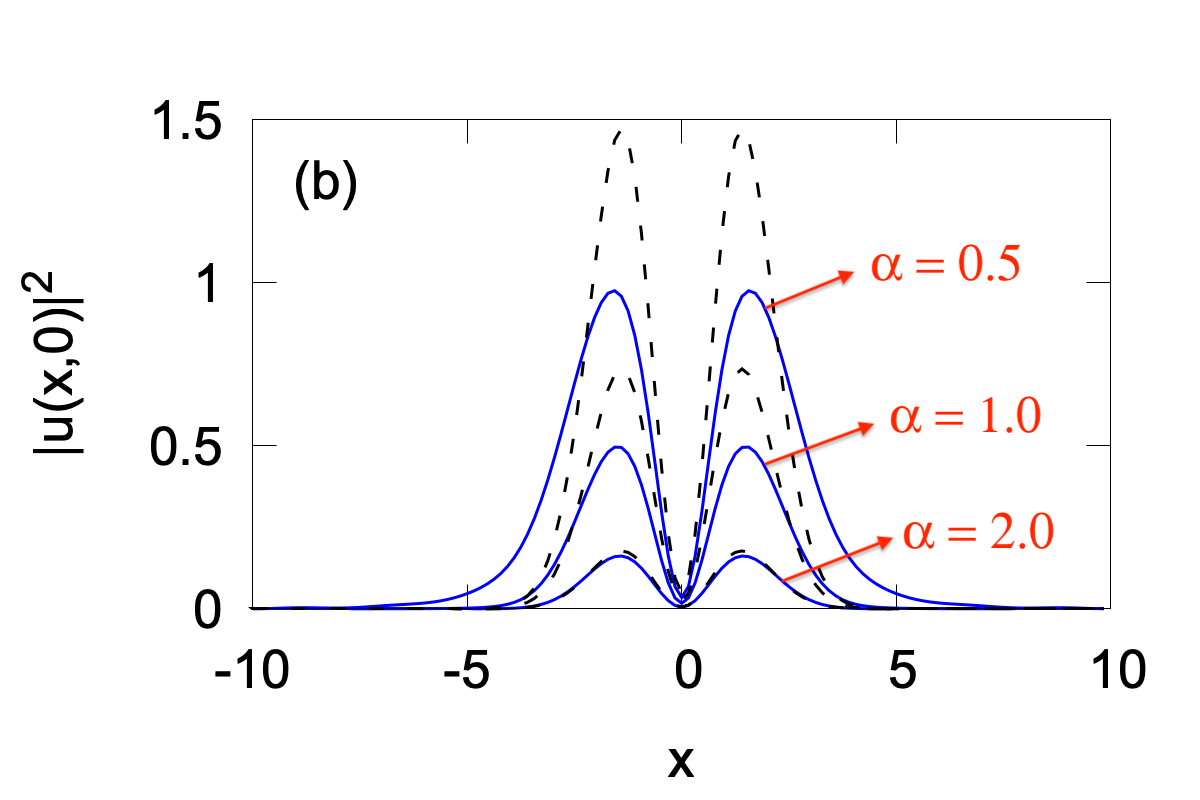}
\caption{The comparison between cross sections (drawn through $y=0$) of the
VA solutions and their numerically found counterparts (dashed black and
solid blue lines, respectively) for different values of the loss parameter $%
\protect\alpha $ in Eq. (\protect\ref{LLE}), taken from set (\protect\ref%
{alpha}). Panels (a) and (b) pertain to the self-focusing ($\protect\sigma %
=+1$) and defocusing ($\protect\sigma =-1$) signs of the cubic nonlinearity,
respectively. The other parameters in Eqs. (\protect\ref{LLE}) and ( \protect
\ref{f}) are fixed as $g=0$, $\protect\eta =1$, $f_{0}=1$, $W=2$, and $S=1$.}
\label{F2}
\end{figure}

In Fig. \ref{F2}(a) we display the cross-section (drawn through $y=0$) of
the variational and numerical solutions for the stable vortex solitons
obtained with $\alpha =0.5$, $1.0$, and $2.0$, while the other parameter are
fixed as $\sigma =+1$ (the self-focusing cubic nonlinearity), $g=0$, $%
f_{0}=1 $, $W=2$, and $S=1$. The accuracy of the VA-predicted solutions
presented in Fig. \ref{F2}(a) is characterized by the relative power
difference from their numerically found counterparts, which is $5.1\%$, $%
3.1\%$, and $0.5\%$ for
\begin{equation}
\alpha =0.5,1.0,\mathrm{~}2.0,  \label{alpha}
\end{equation}%
respectively. Thus, the VA accuracy improves with the increase of $\alpha $.

Similar results for the self-defocusing nonlinearity, $\sigma =-1$, are
presented in Fig. \ref{F2}(b), which shows an essentially larger discrepancy
between the VA and numerical solutions, \textit{viz}., $18.9\%$, $17.4\%$,
and $3.4\%$ for the same set (\ref{alpha}) of values of the loss parameter,
other coefficients being the same as in Fig. \ref{F2}(a). The larger
discrepancy is explained by the fact that localized (bright-soliton) modes
are not naturally maintained by the self-defocusing, hence the ansatz (\ref%
{U}), which is natural for the self-trapped solitons in the case of the
self-focusing, is not accurate enough for $\sigma =-1$. In the same vein, it
is natural that, in the latter case, the discrepancy is more salient for
stronger nonlinearity, i.e., smaller $\alpha $, which makes the respective
amplitude higher.

\begin{figure}[tb]
\centering \includegraphics[width=0.8\columnwidth]{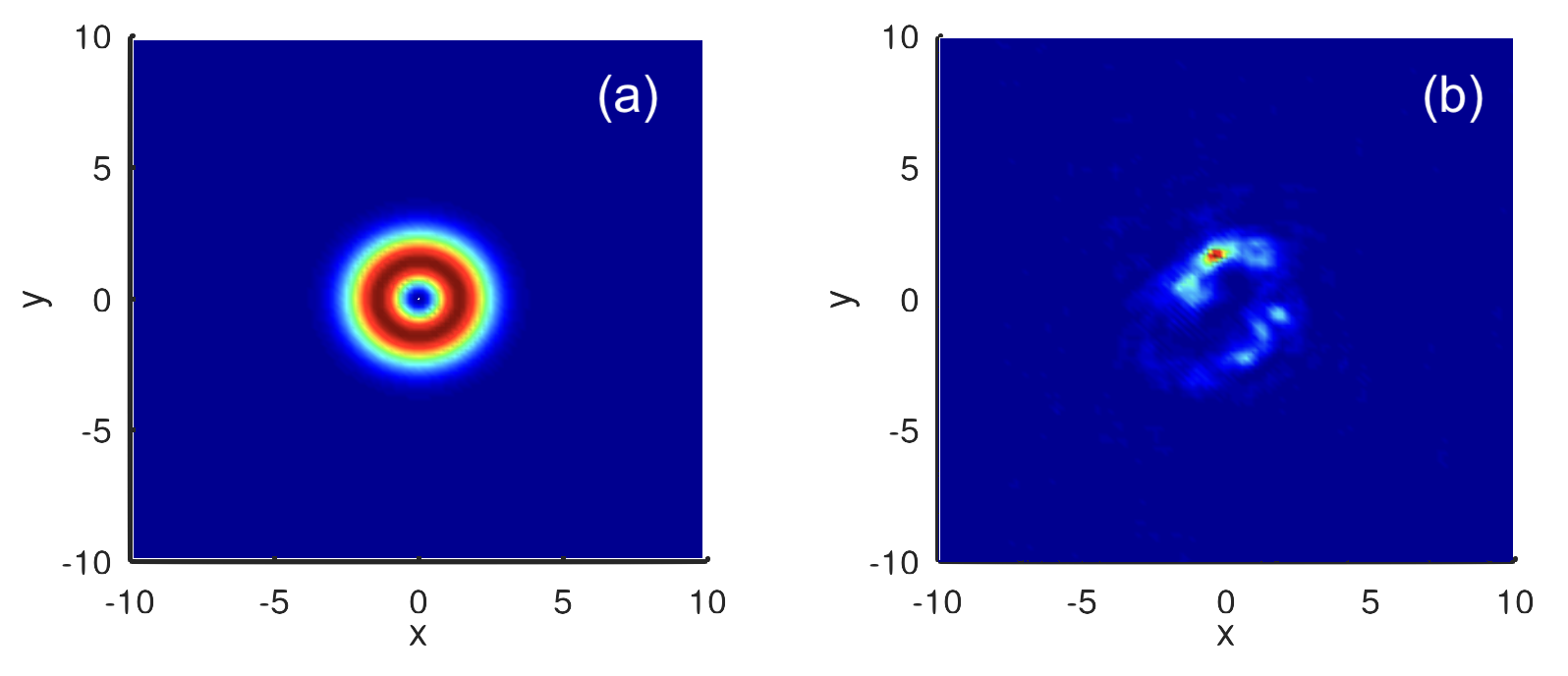}
\caption{Profiles of $|u|^{2}$ produced by VA (a) and numerical solution at $%
t=100$ (b) in the case of the self-focusing ($\protect\sigma =+1$) for $%
\protect\alpha =0.2$. Other parameter are the same as in Fig. \protect\ref%
{F2}(a).}
\label{F3}
\end{figure}

There is a critical value of $\alpha $ below which the vortex solitons are
unstable. As an example, Fig. \ref{F3} shows the VA-predicted and
numerically produced solutions for $\alpha =0.2$ and $\sigma =+1$ (the
self-focusing nonlinearity). The observed picture may be understood as a
result of the above-mentioned azimuthal MI which breaks the axial symmetry
of the vortex soliton. More examples of the instability of this type are
displayed below. For the values of other parameters fixed as in Fig. \ref{F3}%
, the instability boundary is $\alpha _{\mathrm{crit}}\approx 0.35$. The
stabilizing effect of the loss at $\alpha >\alpha _{\mathrm{crit}}$ is a
natural feature. On the other hand, the increase of $\alpha $ leads to
decrease of the soliton's amplitude, as seen in Fig. \ref{F2}.

In the case of the self-defocusing ($\sigma =-1$), all the numerically found
vortex modes are stable, at least, at $\alpha \geq 0.1$, although the
discrepancy in the values of the power between these solutions and their VA
counterparts is very large at small $\alpha $, exceeding $75\%$ at $\alpha
=0.1$. As mentioned above, the growing discrepancy is explained by the
increase of the soliton's amplitude with the decrease of $\alpha $. At still
smaller values of $\alpha $, the relaxation of the evolving numerical
solution toward the stationary state is very slow, which makes it difficult
to identify the stability.

\subsection{Variation of the pump's vorticity $S$}

To analyze effects of the winding number (vorticity) $S$, we here fix $g=0$
(the pure cubic nonlinearity) and set $\alpha =1$, $f_{0}=1$, $W=2$ in Eqs. (%
\ref{LLE}) and (\ref{f}). In the self-focusing case ($\sigma =+1$), the
numerically produced solutions are stable for $S=1$ and $2$, and unstable
for $S\geq 3$. In the former case, the power difference between the VA and
numerical solutions is $3.1\%$ and $4.9\%$ for $S=1$ and $2$, respectively,
i.e., the VA remains a relatively accurate approximation in this case.

In the self-defocusing case ($\sigma =-1$), {considering the same values of
the other parameters as used above,} the numerical solution produces stable
vortex solitons {at least until $S=5$.} For the same reason as mentioned
above, the accuracy of VA is much lower for $\sigma =-1$ than for the
self-focusing case ($\sigma =+1$), with the respective discrepancies in the
power values being $17.4\%$, $6.7\%$, $24.1\%$, {$29.0\%$, and $29.6\%$} for
$S=1$, $2$, $3$, {$4$, and $5$,} respectively.

For the cogent verification of the stability of the localized vortices in
the case of the self-defocusing, we have also checked it for smallest value
of the loss parameter considered in this work, \textit{viz}., $\alpha =0.1$,
again for $S=1$, $2$, $3$, {$4$, and $5$, and the above-mentioned values of
the other coefficients, i.e., }$f_{0}=1$ and $W=2$.{\ Naturally, the
discrepancy between the VA and numerical findings is still higher in this
case, being }$75\%$, $72.5\%$, $65.7\%$, $55.2\%$, and $41.1\%$ for $S=1$, $%
2 $, $3$, {$4$, and $5$,} respectively. {The result is illustrated by Fig. %
\ref{FS4_f1_sigma=-1} for a relatively large vorticity, }$S=4${. In
particular, the pattern of }$|u(x,y)|^{2}$ {and the corresponding cross
section, displayed in Figs. \ref{FS4_f1_sigma=-1}(b) and (c), respectively,
exhibit the established vortex structure and background \textquotedblleft
garbage" produced by the evolution.
\begin{figure}[tb]
\centering\includegraphics[width=0.8\columnwidth]{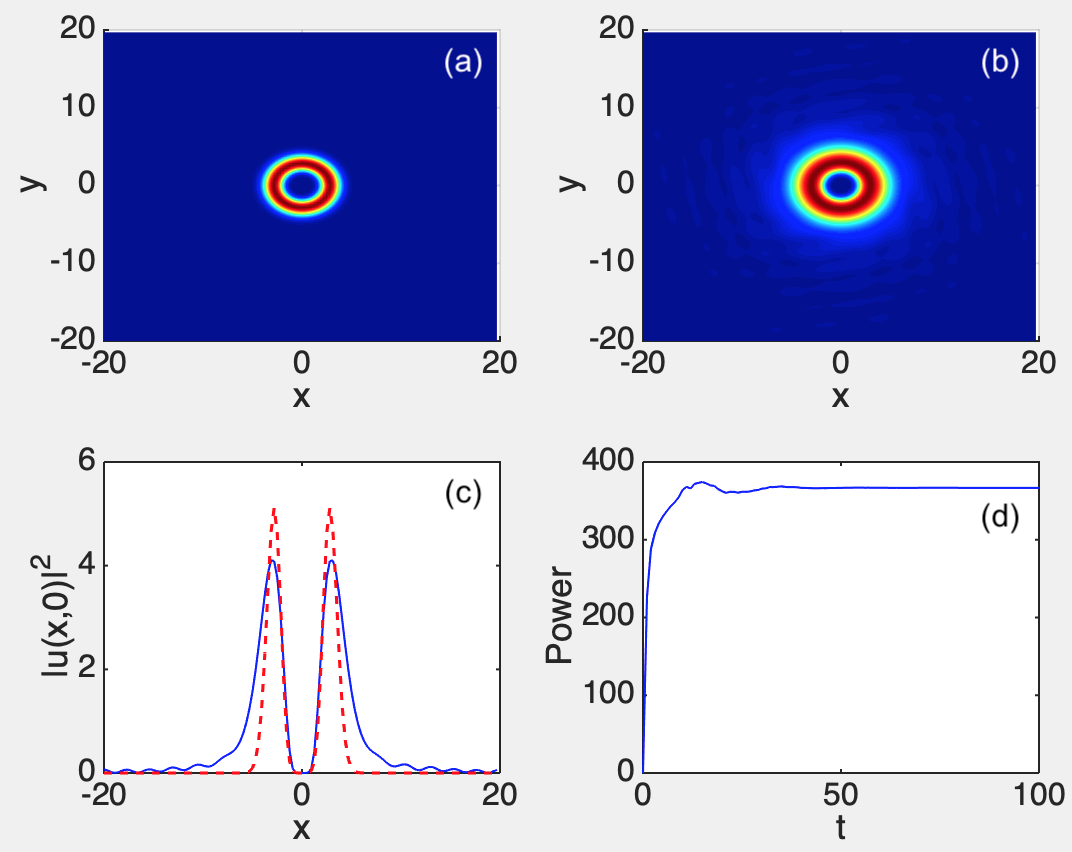}
\caption{(a) The VA-predicted pattern and (b) the corresponding result of
the direct simulation of Eq. (\protect\ref{LLE}) with $\protect\sigma =-1$
and $g=0$ (cubic self-defocusing) at $t=100$, initiated by the zero input at
$t=0$, for $\protect\alpha =0.1$, $f_{0}=1$, $W=2$, and vorticity $S=4$ in
the pump term (\protect\ref{f}). (c) The respective cross sections drawn
through $y=0$. (d) The evolution of the total power $P$ (see Eq. (\protect
\ref{P}) of the numerical solution in the course of the simulation.}
\label{FS4_f1_sigma=-1}
\end{figure}
}

\subsection{Variation of the pump's width $W$}

\begin{figure}[tb]
\centering \includegraphics[width=0.48\columnwidth]{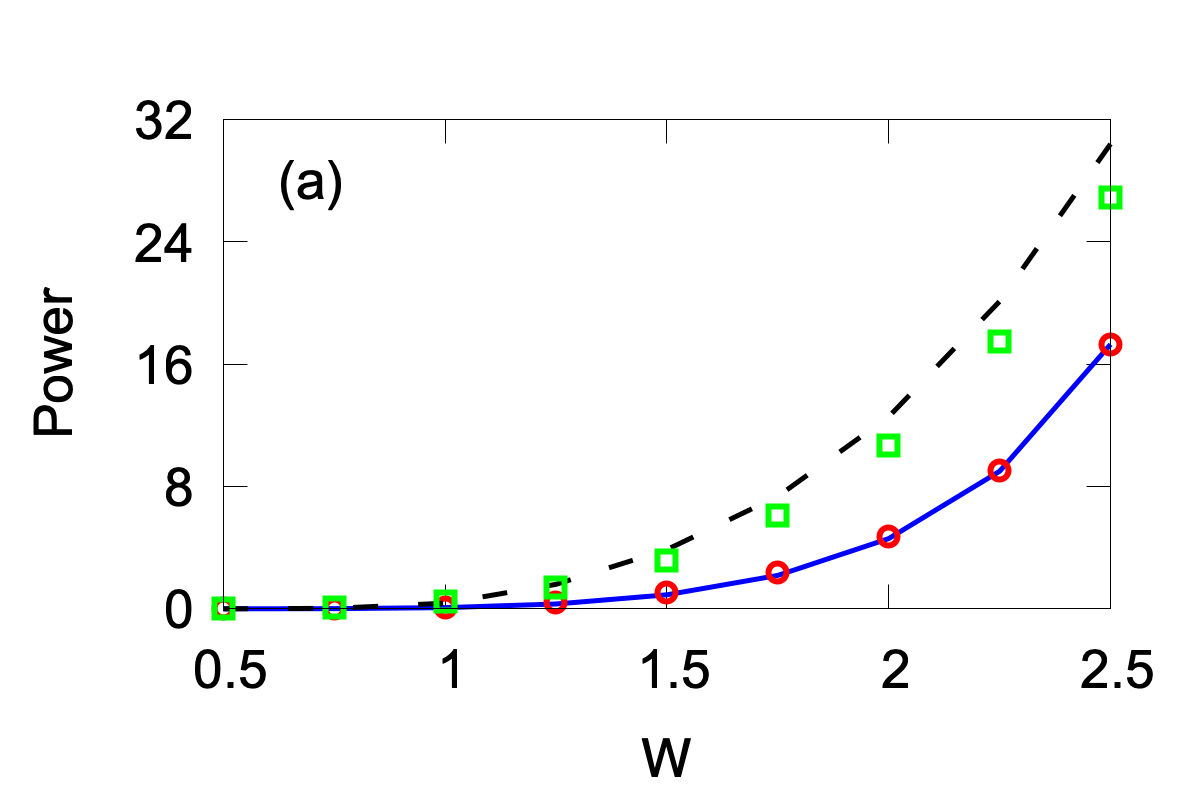} %
\includegraphics[width=0.4\columnwidth]{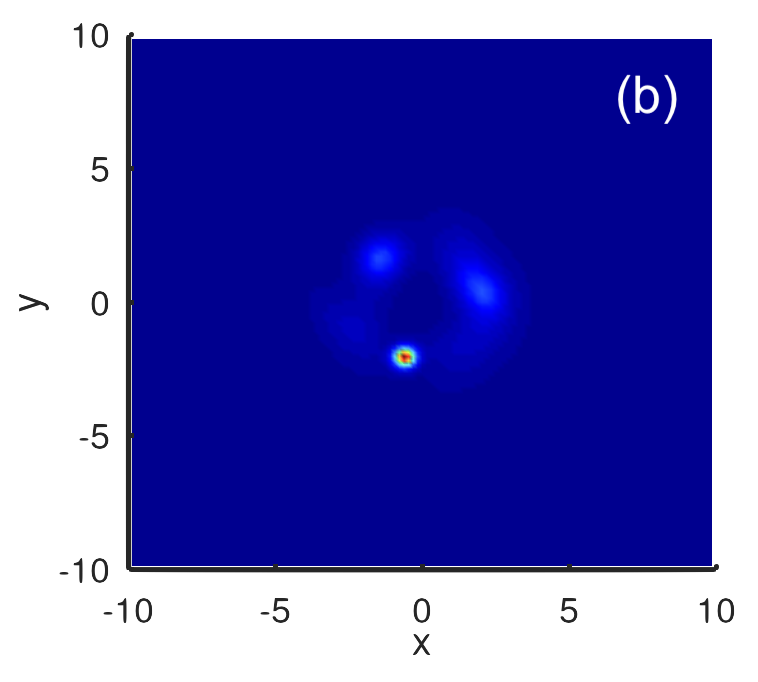}
\caption{(a) The power of the VA-predicted vortex-soliton solutions (solid
blue and dashed black lines, pertaining to the self-focusing, $\protect%
\sigma =+1$, and self-defocusing, $\protect\sigma =-1$, cubic nonlinearitry,
respectively) and their numerically found counterparts (red circles and
green squares pertaining to the self-focusing and self-defocusing
nonlinearitry, respectively) vs. the pump's width $W$. Other parameters are $%
g=0$, $\protect\alpha =1$, $f_{0}=1$, and $S=1$. (b) The profile produced,
at $t=100$, by the numerically generated unstable solution in the case of
the self-focusing, $\protect\sigma =+1$, with $W=2.75$. }
\label{F4}
\end{figure}

To address the effects of the variation of parameter $W$ in Eq. (\ref{f}),
we here fix $g=0$, $\alpha =1$, $f_{0}=1$, and $S=1$. In Fig. \ref{F4}(a),
the power of the VA-predicted and numerically found stable vortex-soliton
solutions is plotted as a function of $W$ for both the self-focusing and
defocusing cases, i.e., $\sigma =+1$ and $\sigma =-1$, respectively. In the
former case, the azimuthal MI sets in at $W\geq 2.7$, see an example in Fig. %
\ref{F4}(b) for $W=2.75$.

In the self-focusing case, $\sigma =+1$, the azimuthal MI for the solitons
with higher vorticities, $S=2$, $3$, $4$, or $5$, sets in at $W\geq 2.1$, $%
1.7$, $1.6$, and $1.4$, respectively. In the self-defocusing case, no
existence/stability boundary was found for the vortex modes with $S=1$, $S=2$%
, and $S=3$ (at least, up to $W=5$). At higher values of the vorticity, the
localized vortices do not exist, in the defocusing case, at $W>3.5$ and $%
W>2.5$, for $S=4$, and $S=5$, respectively.

\subsection{Variation of the pump's strength $f_{0}$}

\begin{figure}[tb]
\centering \includegraphics[width=0.8\columnwidth]{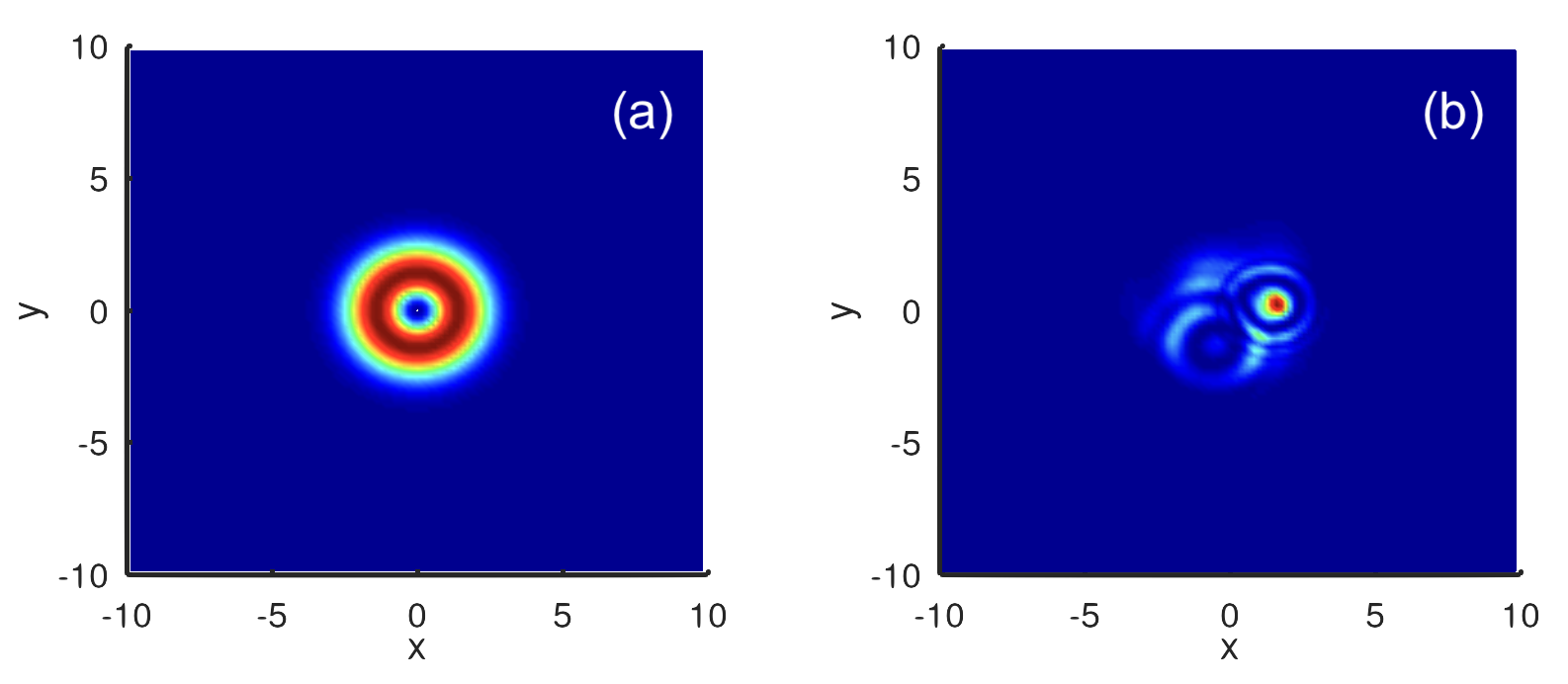}
\caption{(a) The VA-predicted profile of $|u|^{2}$ in the self-focusing
case, with parameters $g=0$, $\protect\sigma =+1$, $\protect\alpha =1$, $%
f_{0}=1.7$, $W=2$, and $S=1$. (b) The unstable solution, produced, at $t=100$%
, by the simulations of Eq. (\protect\ref{LLE}) for the same parameters.}
\label{F5}
\end{figure}

Effects of the variation of $f_{0}$ are reported here, fixing other
parameters as $g=0$, $\alpha =1$, and $W=2$. In the case of the cubic
self-focusing, $\sigma =+1$, the vortex soliton with $S=1$ are subject to MI
at $f_{0}\geq 1.6$. As a typical example, in Fig. \ref{F5} we display the
VA-predicted solution alongside the result of the numerical simulations for $%
f_{0}=1.7$. For higher vorticities, $S=2$, $3$, $4$, and $5$, the azimuthal
instability sets in at $f_{0}\geq 1.1$, $0.6$, $0.3$, and $0.08$,
respectively. Naturally, the narrow vortex rings with large values of $S$
are much more vulnerable to the quasi-one-dimensional azimuthal MI.

\begin{figure}[tb]
\centering \includegraphics[width=0.48\columnwidth]{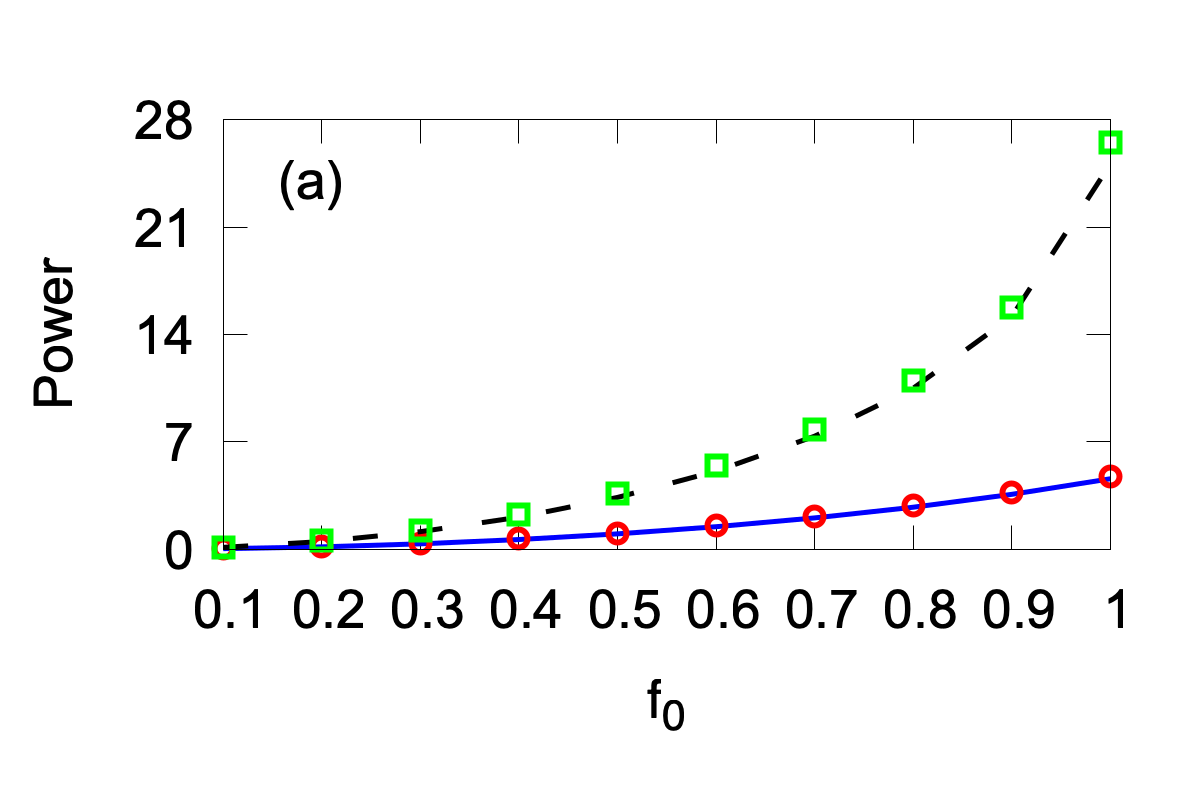} %
\includegraphics[width=0.35\columnwidth]{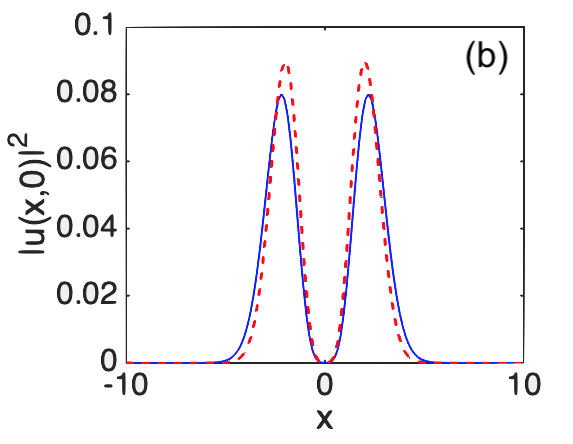}
\caption{(a) The power versus $f_{0}$ for the confined vortex modes in the
self-focusing case ($\protect\sigma =+1$). The VA solutions for $S=1$ and $2$
are shown by solid blue and dashed black lines, respectively. The
corresponding numerical solutions are represented by red circles and green
squares, respectively. Recall that the numerical solutions are stable, in
this case, at $f_{0}<1.6$ and $f_{0}<1.1$, for $S=1$ and $2$, respectively.
(b) The VA-predicted and numerically obtained (the dashed red and solid blue
lines, respectively) profiles of the stable solution with $S=2$ and $%
f_{0}=0.4$, drawn as cross sections through $y=0$ . The other parameters are
$g=0$, $\protect\sigma =+1$, $\protect\alpha =1$, $W=2$.}
\label{F6}
\end{figure}

The power of the vortex solitons with $S=1$ and $2$, as produced by the VA
and numerical solution, is plotted vs. the pump amplitude $f_{0}$ in Fig. %
\ref{F6}(a). As an example, Fig. \ref{F6}(b) showcases an example of the
cross-section profile of the vortex soliton with $S=2$, demonstrating the
reliability of the VA prediction. In the range of $f_{0}\leq 1$, the highest
relative difference in the power between the numerical and variational
solutions cases is $5.5\%$ and $7.5\%$ for $S=1$ and $S=2$, respectively.

It is relevant to mention that the \textquotedblleft
traditional\textquotedblright\ azimuthal instability of vortex-ring solitons
with winding number $S$ demonstrates fission of the original axially
symmetric shape into a set consisting of a large number $N\geq S$ of
symmetrically placed localized fragments \cite{PhysD,book}, while the above
examples, displayed in Figs. \ref{F3}(b), \ref{F4}(b), and \ref{F5}(b),
demonstrate the appearance of a single bright fragment and a
\textquotedblleft garbage cloud\textquotedblright\ distributed along the
original ring. At larger values of $f_{0}$, our simulations produce examples
of the \textquotedblleft clean\textquotedblright\ fragmentation, \textit{viz}%
., with $N=4$ produced by the unstable vortex rings with $S=1$ in Fig. \ref%
{F7}(a), $N=5$ produced by $S=2$ and $3$ in (b) and (d), $N=7$ by $S=2$ and $%
3$ in (c) and (e), and $N=8$ by $S=3$ in (f). These outcomes of the
instability development are observed at the same evolution time $t=100$ as
in Figs. \ref{F3}(b), \ref{F4}(b), and \ref{F5}(b). The gradual increase of
the number of the fragments on \ $S$ is explained by the dependence of the
azimuthal index of the fastest growing eigenmode of the breaking instability
on the underlying winding number $S$, which is a generic property of vortex
solitons \cite{PhysD,book}.

\begin{figure}[tb]
\caption{Examples of the fission of unstable vortex-ring solitons produced
by simulations of the LL equation (\protect\ref{LLE}) with $\protect\sigma %
=+1$ (cubic self-focusing), $g=0$ (no quintic nonlinearity), $\protect\alpha %
=2$, $W=2$, and $\protect\eta =1$. Each plot displays the result of the
numerical simulations at time $t=100$. Values of the initial vorticity and
pump's strength are indicated in panels. }
\label{F7}\centering
\includegraphics[width=1\columnwidth]{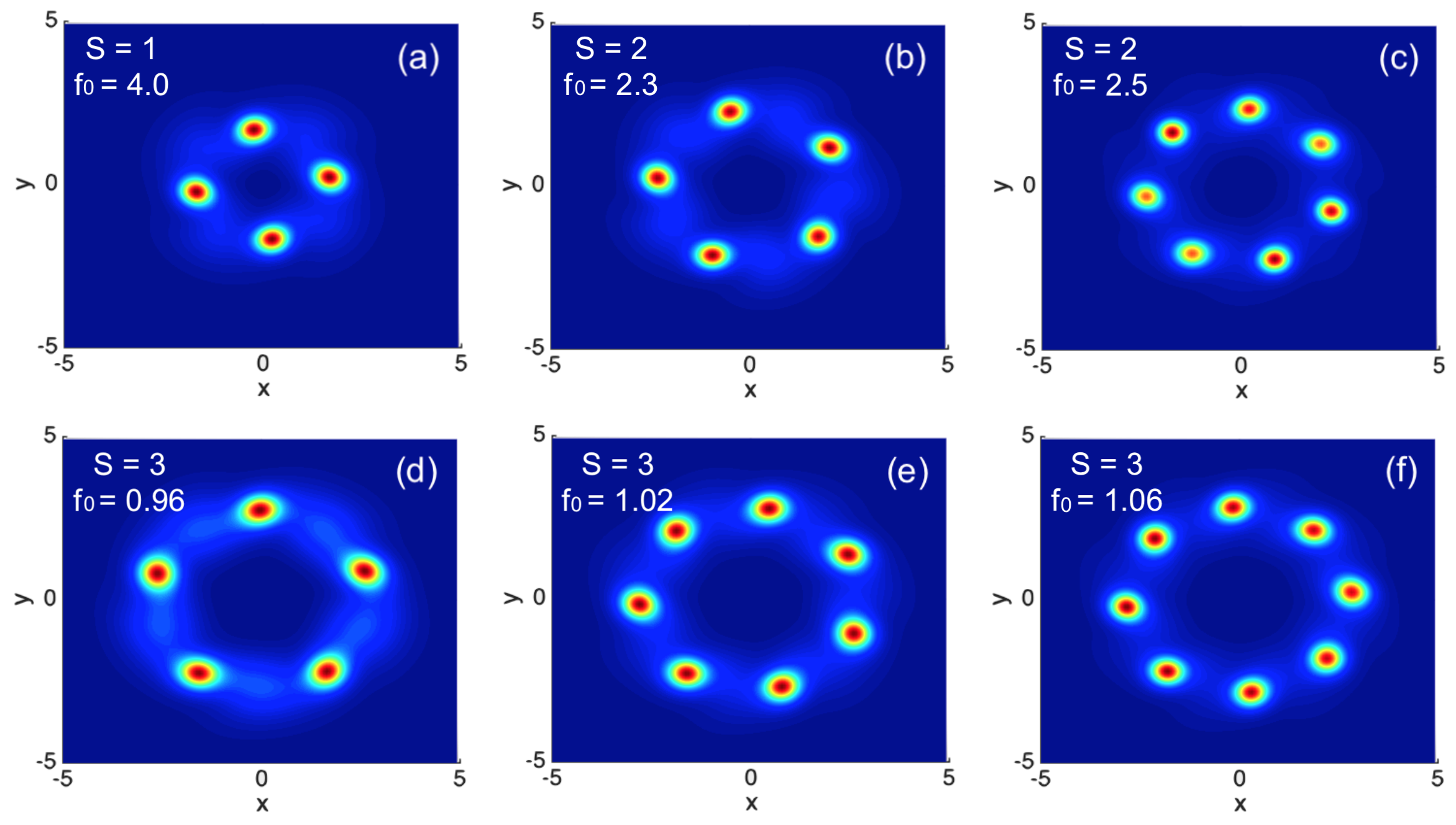}
\end{figure}

\begin{figure}[tb]
\centering \includegraphics[width=0.48\columnwidth]{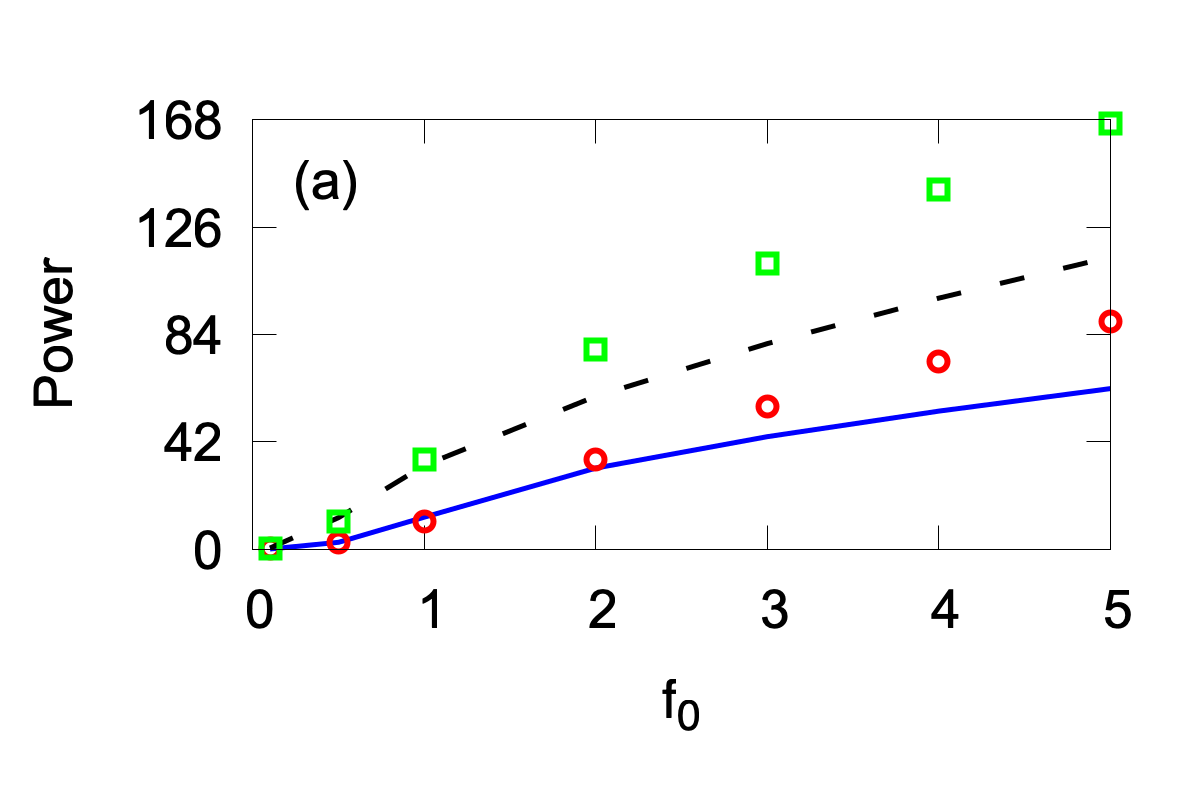} %
\includegraphics[width=0.35\columnwidth]{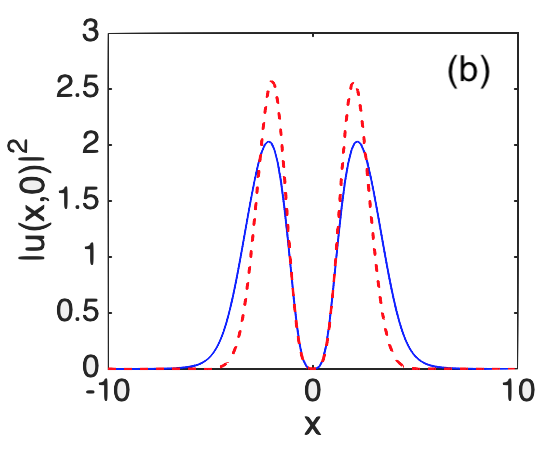}
\caption{(a) The power versus $f_{0}$ for the vortex modes in the
self-defocusing case ($\protect\sigma =-1$). The VA solutions for $S=1$ and $%
2$ are shown by solid blue and dashed black lines, respectively. The
corresponding numerical solutions are presented by red circles and green
squares, respectively. (b) The VA-predicted and numerically obtained (the
dashed red and solid blue lines, respectively) profiles of the solution with
$S=2$ and $f_{0}=2$, drawn as the cross sections through $y=0$ . The other
parameters are $g=0$, $\protect\sigma =-1$, $\protect\alpha =1$, $W=2$.}
\label{F8}
\end{figure}

In self-defocusing case, $\sigma =-1$, a summary of the results produced by
the VA and numerical solution for the stable vortex solitons with $S=1$ and $%
2$, in the form of the dependence of their power on $f_{0}$, is produced in
Fig. \ref{F8}(a) (cf. Fig. \ref{F6}(a) for $\sigma =+1$). Naturally, the
VA-numerical discrepancy increases with the growth of the pump's strength, $%
f_{0}$, see an example in Fig. \ref{F8}(b). Unlike the case of $\sigma =+1$,
in the case of the self-defocusing the vortex modes with $S\leq 5$ remain
stable, at least, up to $f_{0}=5$ (here, we do not consider the case of $S>5$%
).

\subsection{Influence of the quintic coefficient $g$}

In the above analysis, the quintic term was dropped in the LL equation (\ref%
{LLE}), setting $g=0$. To examine the impact of this term, we first address
the case shown above in Fig. \ref{F3}, which demonstrated that the vortex
soliton with $S=1$, as a solution to Eqs. (\ref{LLE}) and (\ref{f}) with $g=0
$, $\sigma =+1$ and $f_{0}=1$, $W=2$, was unstable if the loss parameter
fell below the critical value, $\alpha =0.35$. We have found that, adding to
Eq. (\ref{LLE}) the quintic term with either $g=-1$ or $g=+1$ (the
self-focusing or defocusing quintic nonlinearity, respectively) leads to the
\emph{stabilization} of the vortex mode displayed in Fig. \ref{F3}, which
was unstable in the absence of the quintic term. The stabilization of the
soliton by the quintic self-defocusing is a natural fact. More surprising is
the possibility to provide the stabilization by the self-focusing quintic
nonlinearity because, in most cases, the inclusion of such a term gives rise
to the supercritical collapse in 2D, making all solitons strongly unstable
\cite{PhysD,book}. However, it is concluded from the stability charts
displayed below in Fig. \ref{F11} that the stabilizing effect of the quintic
self-focusing occurs only at moderately small powers, for which the the
quintic term is not a clearly dominant one. In the general case, solitons
stability regions naturally shrink in Figs. \ref{F11} under the action of
the quintic self-focusing quintic term, with $g<0$.

\begin{figure}[tb]
\centering \includegraphics[width=0.4\columnwidth]{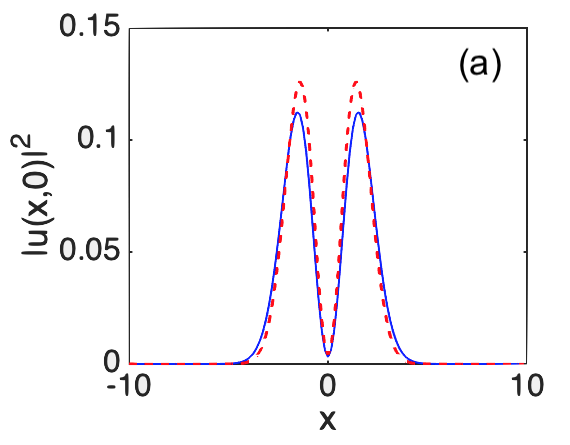} %
\includegraphics[width=0.4\columnwidth]{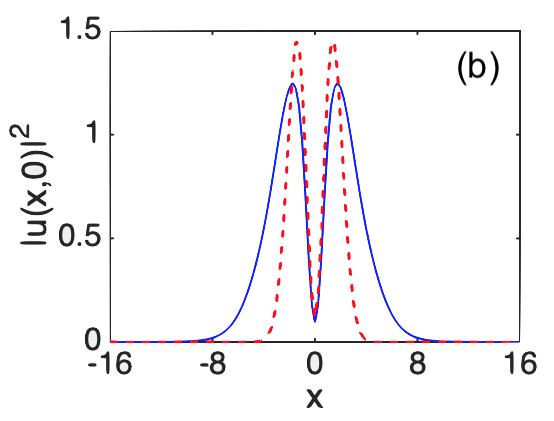}
\caption{Cross-section profiles (drawn through $y=0$) for stable vortex
solitns with $S=1$, produced by Eq. (\protect\ref{LLE}) with the quintic
self-focusing, $g=-1$ (a) or defocusing, $g=+1$ (b) term. In both cases, the
cubic self-focusing cubic term, with $\protect\sigma =+1$, is present. The
numerically found profiles and their VA-produced counterparts are displayed,
respectively, by the solid blue and dashed lines. Other parameter are $%
\protect\alpha =0.2$, $\protect\eta =1$, and $f_{0}=1$, $W=2$.}
\label{F9}
\end{figure}

In the presence of the quintic term, the comparison of the numerically found
stabilized vortex-soliton profiles with their VA counterparts, whose
parameters are produced by a numerical solution of Eqs. (\ref{EL1}) and (\ref%
{EL2}), is presented in Fig. \ref{F9}. Similar to the results for the LL
equation with the cubic-only nonlinearity, the VA is essentially more
accurate in the case of the self-focusing sign of the quintic term ($g<0$)
than in the opposite case, $g>0$. In particular, in the case shown in Fig. %
\ref{F9}, the power-measured discrepancy for $g=-1$ and $+1$ is,
respectively, $4\%$ and $64\%$. An explanation for this observation is
provided by the fact that the soliton's amplitude is much higher in the
latter case.

\begin{figure}[tb]
\centering \includegraphics[width=0.4\columnwidth]{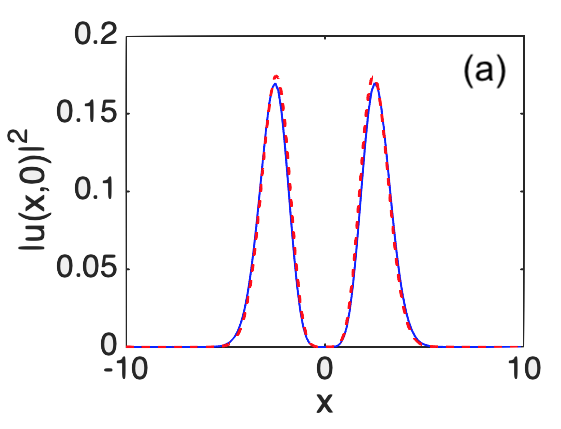} %
\includegraphics[width=0.4\columnwidth]{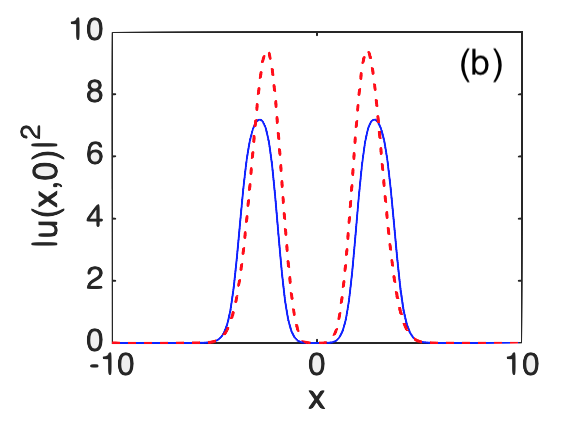}
\caption{Cross-section profiles (drawn through $y=0$) of vortex solitons
with $S=3$, stabilized by the self-focusing (a) or defocusing (b) quintic
tterm in Eq. (\protect\ref{LLE}), with the respective coefficient $g=-6$ or $%
g=0.1$, other parameters being $\protect\sigma =-1$ (cubic self-focusing), $%
\protect\alpha =1$, $\protect\eta =1$, and $f_{0}=1$, $W=2$ in Eq.( \protect
\ref{f}). The numerically found solutions and their VA-predicted
counterparts are plotted by the solid blue and dashed red lines,
respectively.}
\label{F10}
\end{figure}

Another noteworthy finding is the stabilization of higher-vorticity solitons
by the quintic term. For instance, it was shown above that, for parameters $%
\alpha =1$, $\eta =1$, $f_{0}=1$, $W=2$, and $\sigma =+1$ (the cubic
self-focusing) all vortex solitons with $S\geq 3$, produced by Eq. (\ref{LLE}%
), were unstable. Now, we demonstrate that the soliton with $S=3$ is
stabilized by adding the self-defocusing quintic term with a small
coefficient, just $g=0.1$, see Fig. \ref{F10}(b). As a counter-intuitive
effect, the stabilization of the same soliton by the self-focusing quintic
term is possible too, but the necessary coefficient is large, $g=-6$, see
Fig. \ref{F10}(a) (recall that $g=-1$ is sufficient for the stabilization of
the vortex soliton with $S=1$ and $\alpha =0.2$ in Fig. \ref{F9}(a)).
Nevertheless, similar to what is said above, the set of Figs. \ref{F11}%
(c,f,i) demonstrates the natural shrinkage of the stability area under the
action of the quintic self-focusing. For the stabilized vortex modes shown
in Fig. \ref{F10}(a), in the case when both the cubic and quintic terms are
self-focusing, the relative power-measured discrepancy between the numerical
and VA-predicted solutions is very small, $\approx 0.7\%$, while in the
presence of the weak quintic self-defocusing in Fig. \ref{F10}(b) the
discrepancy is $6.6\%$.

\subsection{Stability charts in the parameter space}

\begin{figure}[t]
\centering \includegraphics[width=0.9\textwidth]{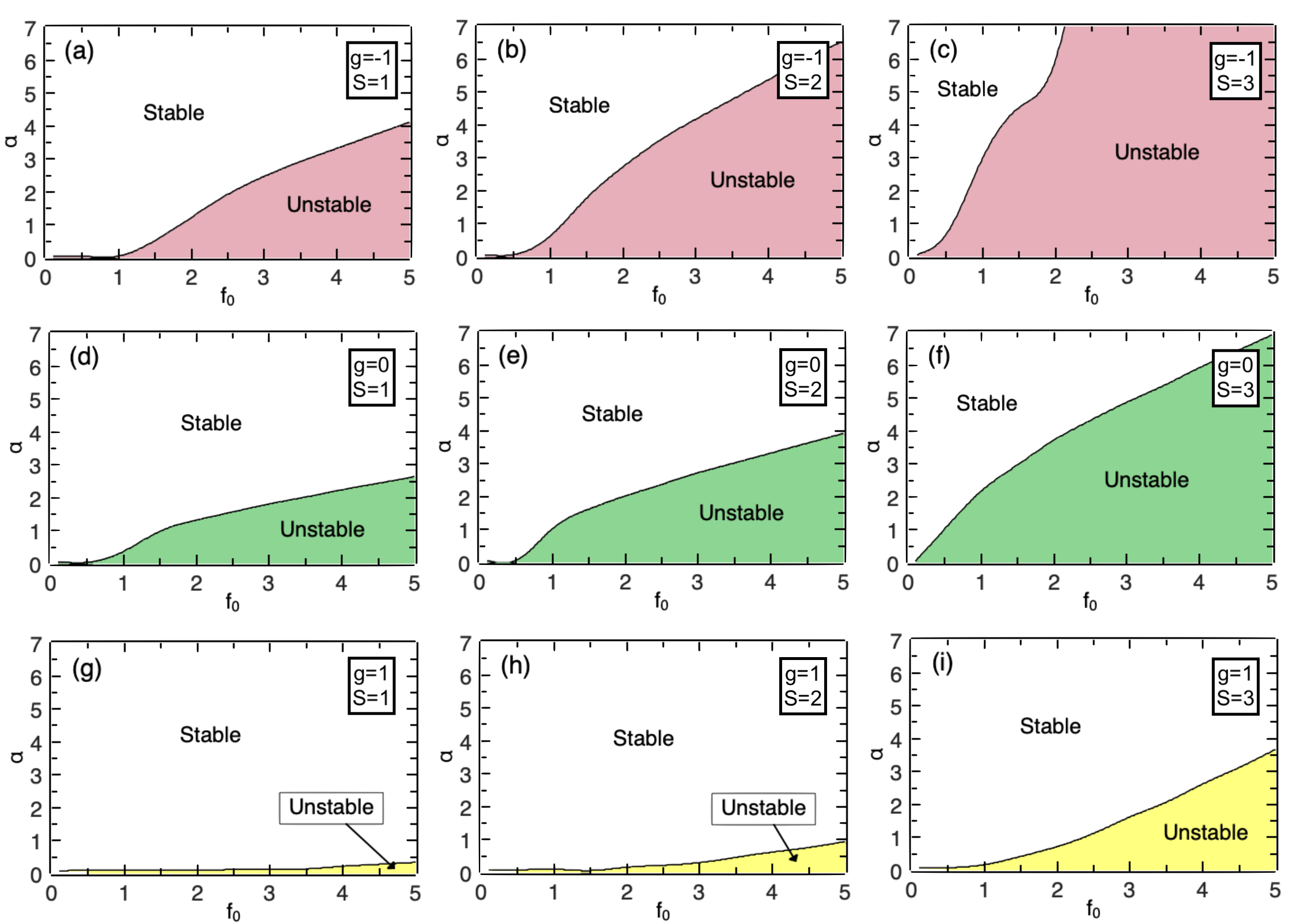}
\caption{Stability areas for\ families of the vortex solitons with winding
numbers $S=1$, $2,$ and $3$, in the plain of the loss coefficient ($\protect%
\alpha $) and amplitude of the pump beam ($f_{0}$), for three different
values of the quintic coefficient, $g=-1,0,+1$ (recall that $g<0$ and $g>0$
correspond, respectively, to the self-focusing and defocusing). Other
parameters of Eqs. (\protect\ref{LLE}) and (\protect\ref{f}) are $\protect%
\sigma =+1$ (the self-focusing cubic term), $\protect\eta =1$, and $W=2$.}
\label{F11}
\end{figure}

The numerical results produced in this work are summarized in the form of
stability areas plotted in Fig. \ref{F11} in the parameter plane $\left(
f_{0},\alpha \right) $, for the vortex-soliton families with winding numbers
$S=1$, $2$, $3$, and three values of the quintic coefficient, $g=-1$, $0$, $%
+1$, while the cubic term is self-focusing, $\sigma =+1$, and the width of
the pump beam is fixed, $W=2$. In addition to that, stability charts
corresponding to the combination of the cubic self-defocusing ($\sigma =-1$)
and quintic focusing ($g=-1$), also for $S=1,2,3$, are plotted in Fig. \ref%
{F12}.

\begin{figure}[tbph]
\centering
\includegraphics[width=0.9\textwidth]{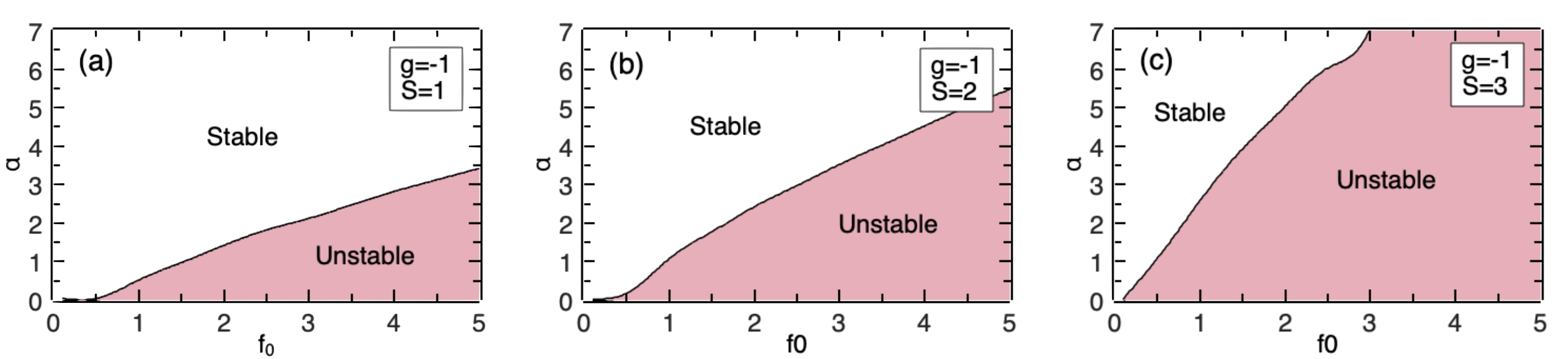}
\caption{The same as in Figs. \protect\ref{F11}(a-c), but for $\protect%
\sigma =-1$ and $g=-1$ (the cubic self-defocusing and quintic focusing
terms).}
\label{F12}
\end{figure}

The choice of the parameter plane $\left( f_{0},\alpha \right) $ in the
stability diagrams displayed in Fig. \ref{F11} is relevant, as the strength
of the pump beam, $f_{0}$, and loss coefficient, $\alpha $,are amenable to
accurate adjustment in the experiment (in particular, $\alpha $ may be tuned
by partially compensating the background loss of the optical cavity by a
spatially uniform pump, taken separately from the confined pump beam). As
seen in all panels of Figs. \ref{F11} and \ref{F12}, the increase of $\alpha
$ naturally provides effective stabilization of the vortex modes, while none
of them may be stable at $\alpha =0$, in agreement with the known properties
of vortex-soliton solutions of the 2D nonlinear Schrödinger equation with
the cubic and/or cubic-quintic nonlinearity \cite{PhysD,book}. The apparent
destabilization of the vortices with the increase of the pump's amplitude $%
f_{0}$ is explained by the ensuing enhancement of the destabilizing
nonlinearity. Other natural features exhibited by Figs. \ref{F11} are the
general stabilizing/destabilizing effect of the quintic
self-defocusing/focusing (as discussed above), and expansion of the
splitting-instability area with the increase of $S$ (the latter feature is
also exhibited by Fig. \ref{F12}). The latter finding is natural too, as
larger $S$ makes the ring-shaped mode closer to the quasi-1D shape (see, in
particular, Fig. \ref{F7}), which facilitates the onset of the
above-mentioned azimuthal MI (modulational instability).

Thus, the inference is that the instability mode which determines the
boundary of the stability areas in Figs. \ref{F11} and \ref{F12} is the
breaking of the axial symmetry of the vortex rings by azimuthal
perturbations, as shown above, in particular, in Figs. \ref{F3}(b), \ref{F4}%
(b), and \ref{F5}(b). The destabilization through the spontaneous splitting
of the rings into symmetric sets of fragments (see Fig. \ref{F7}) occurs
deeply inside the instability area, i.e., at larger values of $f_{0}$.

\section{Conclusion \label{sec:Conclusion}}

We have introduced the two-dimensional LL (Lugiato-Lefever) equation
including the self-focusing or defocusing cubic or cubic-quintic
nonlinearity and the confined pump with embedded vorticity (winding number),
$S\leq 5$. Stable states in the form of vortex solitons (rings) for these
values of $S$ are obtained, in parallel, in the semi-analytical form by
means of the VA (variational approximation) and numerically, by means of
systematic simulations of the LL equation starting from the zero input. The
VA provides much more accurate results in the case of the self-focusing
nonlinearity than for the defocusing system. Stability areas of the vortex
solitons with $S=1,2,3$ are identified in the plane of experimentally
relevant parameters, \textit{viz}., the pump amplitude and loss coefficient,
for the self-focusing and defocusing signs of the cubic and quintic terms.
Stability boundaries for the vortex rings are determined by the onset of the
azimuthal instability which breaks their axial symmetry. These findings
suggest new possibilities for the design of tightly confined robust optical
modes, such as vortex pixels.

As an extension of this work, it may be interesting to construct solutions
pinned to a symmetric pair of pump beams with or without intrinsic
vorticity. In this context, it is possible to consider the beam pair with
identical or opposite vorticities. In the case of the self-focusing sign of
the nonlinearity, one may expect onset of spontaneous breaking of the
symmetry in the dual-pump configuration. Results for this setup will be
reported elsewhere.

\section*{Acknowledgments}

We thank Prof. Branko Dragovich for invitation to submit the paper to the
Special Issue of Symmetry on the topic of \textquotedblleft Selected Papers
on Nonlinear Dynamics".

The work of S.K. and B.A.M. is supported, in part, by the Israel Science
Foundation through grant No. 1695/22. W.B.C. acknowledges the financial
support of the Brazilian agency CNPq (grant \#306105/2022-5). This work was
also performed as a part of program \#465469/2014-0 of the Brazilian
National Institute of Science and Technology (INCT) for Quantum Information.

\end{document}